\documentclass[%
 aip,
 jmp,%
 amsmath,amssymb,
 reprint,%
]{revtex4-1}

\usepackage{graphicx}
\usepackage{dcolumn}
\usepackage{bm}
\usepackage[toc,page]{appendix}
\usepackage{color}

\begin{document}
\title[Variable high gradient permanent magnet quadrupole]{Variable high gradient permanent magnet quadrupole (QUAPEVA)}

\author{F . Marteau}
\affiliation{Synchrotron-SOLEIL, Saint-Aubin, Gif-sur-Yvette 91192, FRANCE.}

\author{P. N'gotta}
\affiliation{Synchrotron-SOLEIL, Saint-Aubin, Gif-sur-Yvette 91192, FRANCE.}
 
 \author{C. Benabderrahmane}
\affiliation{ESRF, 71 Avenue des martyrs, Grenoble 38000, FRANCE.}

\author{A. Ghaith}
\affiliation{Synchrotron-SOLEIL, Saint-Aubin, Gif-sur-Yvette 91192, FRANCE.}
\affiliation{ Universit\'e Paris-Saclay,  FRANCE.}

\author{M. Vall\'eau}
\affiliation{Synchrotron-SOLEIL, Saint-Aubin, Gif-sur-Yvette 91192, FRANCE.}

\author{A. Loulergue}
\affiliation{Synchrotron-SOLEIL, Saint-Aubin, Gif-sur-Yvette 91192, FRANCE.}

\author{J. V\'et\'eran}
\affiliation{Synchrotron-SOLEIL, Saint-Aubin, Gif-sur-Yvette 91192, FRANCE.}

\author{M. Sebdaoui}
\affiliation{Synchrotron-SOLEIL, Saint-Aubin, Gif-sur-Yvette 91192, FRANCE.}

\author{T. Andr\'e}
\affiliation{Synchrotron-SOLEIL, Saint-Aubin, Gif-sur-Yvette 91192, FRANCE.}

 \author{G. Le Bec}
\affiliation{ESRF, 71 Avenue des martyrs, Grenoble 38000, FRANCE.}

 \author{J. Chavanne}
\affiliation{ESRF, 71 Avenue des martyrs, Grenoble 38000, FRANCE.}

 \author{C. Vallerand}
\affiliation{LAL,  Centre scientifique d'Orsay, Bt 200, BP 34, 91898, FRANCE.}

\author{D. Oumbarek}
\affiliation{Synchrotron-SOLEIL, Saint-Aubin, Gif-sur-Yvette 91192, FRANCE.}

\author{O. Cosson}
\affiliation{SigmaPhi, Rue des Fr{\`e}res Montgolfier, 56000 Vannes, FRANCE}

\author{F. Forest}
\affiliation{SigmaPhi, Rue des Fr{\`e}res Montgolfier, 56000 Vannes, FRANCE}

\author{ P. Jivkov}
\affiliation{SigmaPhi, Rue des Fr{\`e}res Montgolfier, 56000 Vannes, FRANCE}

\author{ J. L. Lancelot}
\affiliation{SigmaPhi, Rue des Fr{\`e}res Montgolfier, 56000 Vannes, FRANCE}

\author{M.E. Couprie}
\affiliation{Synchrotron-SOLEIL, Saint-Aubin, Gif-sur-Yvette 91192, FRANCE.}
\affiliation{ Universit\'e Paris-Saclay,  FRANCE.}

\date{\today}

\begin{abstract}

High gradient quadrupoles are necessary for different applications such as laser plasma acceleration, colliders, and diffraction limited light sources. Permanent magnet quadrupoles provide a higher field strength and compactness than conventional electro-magnets. An original design of permanent magnet based quadrupole (so-called "QUAPEVA"), composed of a Halbach ring placed in the center with a bore radius of 6 mm and surrounded by four permanent magnet cylinders capable of providing a gradient of 210 T/m, is presented. The design of the QUAPEVAs, including magnetic simulation modeling, and mechanical issues are reported. 
Magnetic measurements of seven systems of different lengths are presented and confirmed the theoretical expectations. The variation of the magnetic center while changing the gradient strength is $\pm$ 10 $\mu m$. 
A triplet of three QUAPEVA magnets are used to focus a beam with large energy spread and high divergence that is generated by Laser Plasma Acceleration source for a free electron laser demonstration.

\end{abstract}

\pacs{41.85.Lc, 75.50.Bb}
\keywords{Quadrupole, permanent magnet}

\maketitle

%


\section{Introduction}
 Accelerator physics and technology have recently seen tremendous developments. For example, colliders aim at beam focus at nanometer size scale for high energy physics applications\cite{balakin1995focusing, OidePhysRevAccelBeams.19.111005}, and thus require strong quadrupolar fields. The domain of synchrotron radiation is actively investigating low emittance storage rings (picometer scale) with multibend achromat optics for getting closer to the diffraction limit and providing a high degree of transverse coherence \cite{cai2012ultimate}, for which high gradient quadrupoles with a small harmonic content is one of the issues. In addition, Laser Plasma Acceleration (LPA) can now generate a GeV beam within a very short accelerating distance (few centimeters), with high peak current $\sim$10 kA, but the high divergence (few mrads) and large energy spread (few percent) have to be handled \cite{couprie2014towards}.

 All these recent developments have, in common, requirement of high gradient quadrupoles. Permanent Magnet Quadrupoles (PMQs) achieve high gradient with compactness and with the absence of power supplies, letting them to be a solution for future sustainable green society. 
 Several Halbach\cite{halbach1983permanent,halbach1983conceptual} ring based PMQs with fixed gradient were designed and built: 
at CESR \cite{lou1998stability};   
at Kyoto University / SLAC \cite{mihara2004super};
 at the department f$\ddot{u}$r Physik\cite{eichner2007miniature}; 
at ESRF \cite{Ngotta2016hybrid}...
Various original designs were proposed and developed for the permanent magnet quadrupole to provide a variable gradient, such as at SLAC / Fermilab collaboration)\cite{gottschalk2005performance}; at Kyoto U. / SLAC collaboration\cite{mihara2006variable}; at STFC Daresbury Laboratory  CERN collaboration for the CLIC project \cite{shepherd2012novel}...
So far, the PMQ gradient variability range is limited, and multipolar contents are not yet as low as with electromagnetic technology.

In this paper, an original hybrid compact permanent magnet based quadrupole of variable strength (QUAPEVA) \cite{BenabWO2016034490, BenadWOBL14SSOQUA} developed at Synchrotron SOLEIL is presented. 
The QUAPEVA concept provides a large gradient tunability  $50\%$, a gradient reaching 210 $T/m$, and a center stability excursion while varying the gradient of $\pm$10 $\mu m$. Simulation models and the mechanical design are presented. Seven systems of different integrated gradients have been built. Magnetic measurements using three different methods are shown. Finally, three QUAPEVAs have been used for the  COXINEL project\cite{couprie2016application} aiming at demonstrating LPA based Free Electron Laser (FEL) amplification. 

\section{\label{sec:level1}QUAPEVA Concept}

Let's consider the local field $B(x, y, s)$ inside a quadrupole, with x (resp. y) the horizontal (resp.  vertical) direction, and s the longitudinal axis. In case of an infinitely long magnet, the complex induction $B(z)=B_y + iB_x$ with $z=x+iy$, representing the vertical $B_y$ and horizontal $B_x$ components, can be expressed as follows: $B(z)=\sum_{n=1}^{\infty} (B_n+iA_n)\frac{z^{n-1}}{r_{0}}$, where n is the multipolar order, $B_n$ and $A_n$ are the normal and skew multipolar coefficients respectively, $r_0$ the radius for which coefficients are computed or measured. "Normalized" components $a_n$ and $b_n$ are defined as $a_n =10^4.A_n/B_2$ and $b_n = 10^4.B_n/B_2$.
For a perfect normal quadrupole (n=2, $A_2$=0), the complex induction becomes: $B_y+iB_x=B_2\frac{x+iy}{r_0}$. This evolution is the same for the horizontal field along the vertical axis, as for a realistic quadrupole, it contains higher order multipoles resulting from the structure, magnets, or mechanical assembly imperfections. 

\begin{figure}[ht]
\centering
\includegraphics[scale=0.24]{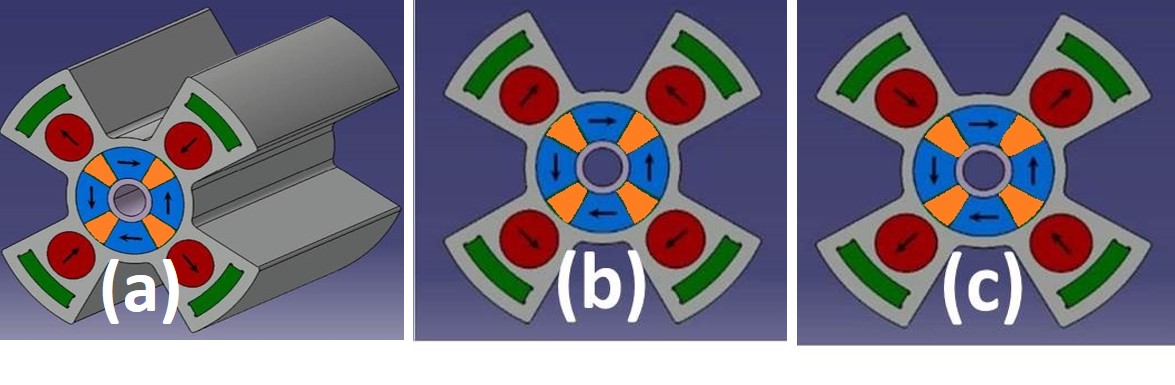}
\caption{Scheme of the QUAPEVA: Permanent magnet blocks (Blue) and rotating cylinders (Red), Vanadium Premendur magnetic plates (Green) and poles (Orange), Aluminum support frame (Grey). (a) maximum, (b) intermediate, and (c) minimum gradient.}
\label{Fig1}
\end{figure}

Fig. \ref{Fig1} presents three particular configurations of the tuning magnets; (a) maximum gradient: tuning magnets easy axis towards the central magnetic poles, (b) intermediate gradient: the tuning magnets are in the reference position, i.e. their easy axis is perpendicular to the central magnetic poles, (c) minimum gradient: tuning magnets easy axis away from the central magnetic poles.

The first quadrupole is a Halbach hybrid structure constituted with four $Nd_2Fe_{14}B$ PMs, and four Iron-Cobalt alloy magnetic poles. The second one is composed of four PM cylinders of radius $a$ with a radial magnetic moment orientation, each connected to a motor, which produces a variable gradient by the rotation around their axis.
Four Fe-Co alloys are placed behind the PM cylinders to shield the magnetic field. The magnetic system is inserted into a dedicated Aluminum support frame in order to maintain the magnetic elements in their positions due to the strong generated magnetic force.

\section{\label{sec:level1}QUAPEVA design}

The QUAPEVA specifications have been defined according to LPA beam transport in the case of COXINEL experiment. QUAPEVAs should be compact with bore radius of 6 mm and adequate to vacuum environment, have magnetic lengths from 26 mm up to 100 mm, magnets quality should ensure high remanence and coercivity, and the design should guaranty a high gradient G$\geqslant$ 100 T/m with a large tunability $\geqslant 30\%$, alongside small harmonic components ($b_{6}/b_{2}\leqslant 3\%$, $b_{10}/b_{2}\leqslant 1\%$). Motors also should be able to handle the magnetic forces induced by the magnetic system.

\begin{figure}[h!]
\centering
\includegraphics[scale=0.35]{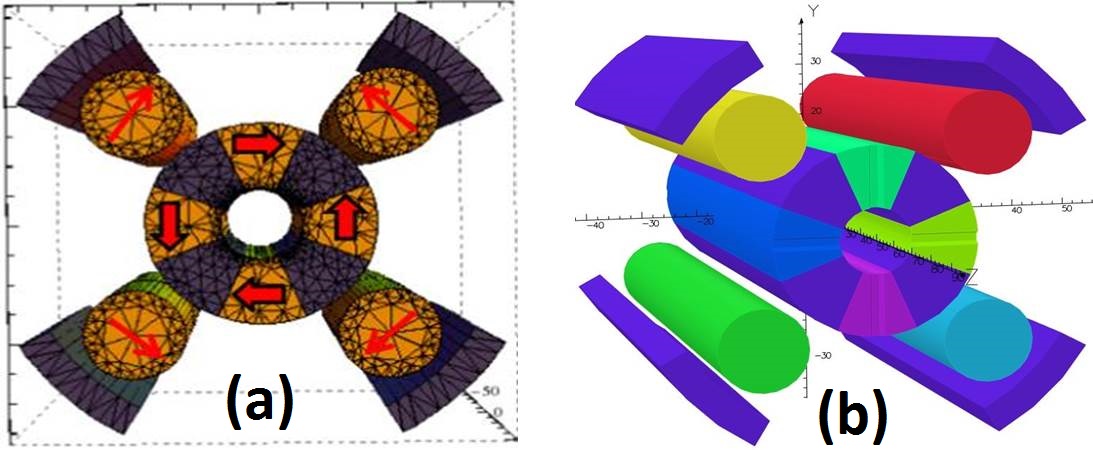}
\caption{(a) RADIA model, (b) TOSCA model.}
\label{Fig2}
\end{figure}

Two numerical tools are used to optimize the geometry and magnetic parameters of the QUAPEVAs: RADIA\cite{RADIAChubarJSR1998} a magnetostatic code based on boundary integral method ( Fig. \ref{Fig2}-a); TOSCA\cite{TOSCA} a finite element magnetostatic code (Fig. \ref{Fig2}-b). The tuning magnets magnetization angles are parameterized in order to simulate the gradient tuning and check the gradient range.

\begin{figure}[h!]
\centering
\includegraphics[scale=0.36]{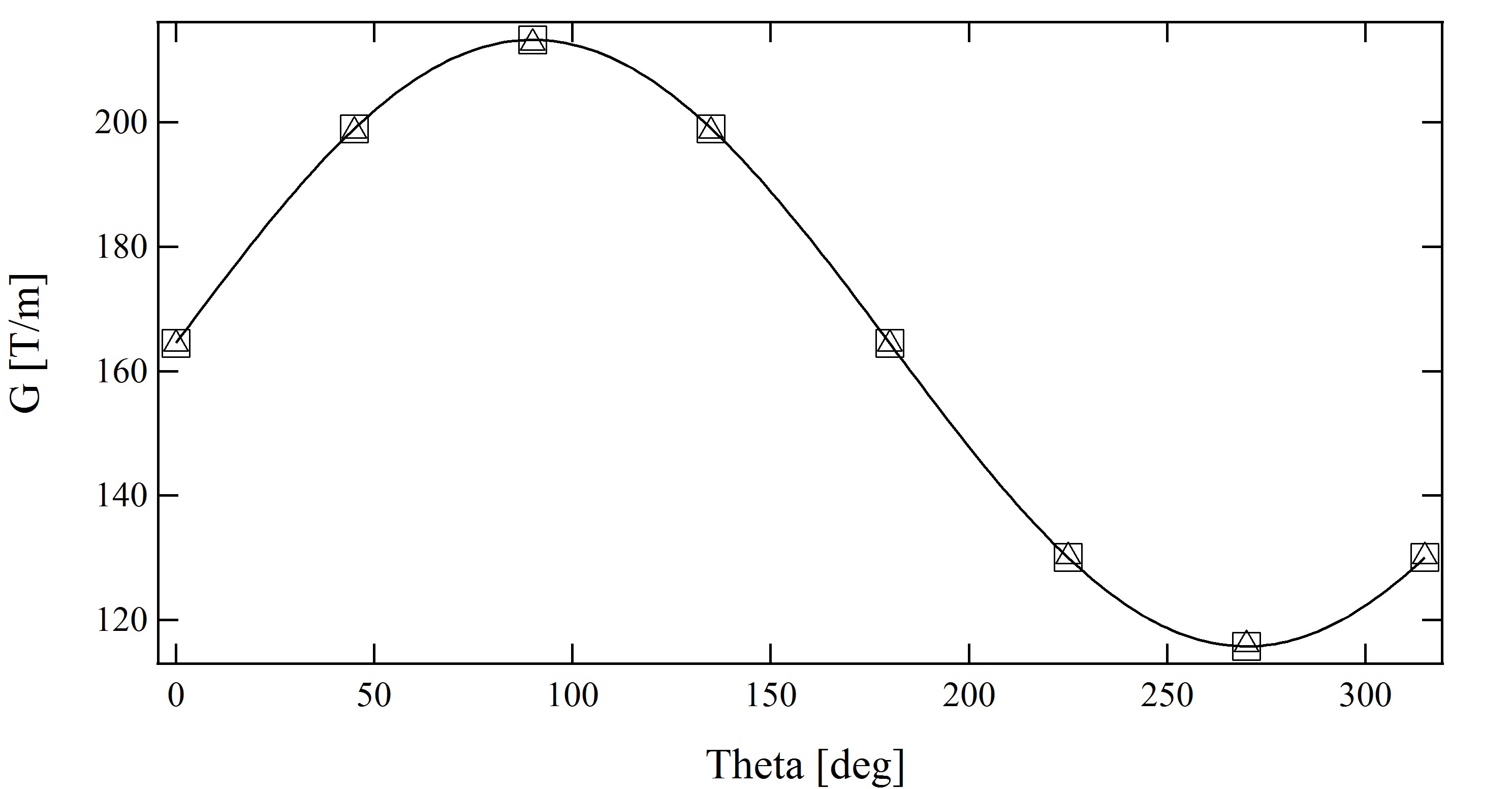}
\caption{Simulations of the prototype gradient evolution versus tuning magnets angle with ($\triangle$) TOSCA and ($\square$) RADIA. (Line) sinus fit.  Remanent field: 1.26 T , coercivity: 1830 kA/m, pole saturation: 2.35 T}
\label{Fig3}
\end{figure}

Fig. \ref{Fig3} shows the simulated gradient evolution with tuning magnet angle. The tuning magnets are rotated by the same angle from their reference position, where the gradient reaches a maximum and a minimum value for a complete rotation. The simulation results of the two models are in good agreement. The evolution is fitted with a sinus function $G(\theta)=G_0 + G_t sin(\theta)$, where $G_0$ is the fixed gradient of the main magnets, $G_t$ the gradient contribution of all the tuning magnets, and $\theta$ their corresponding angle. The gradient variation from peak to peak is $\sim$100 T/m and the maximum gradient reaches $\sim$210 T/m with the prototype one (l= 100 mm). The simulated field multipole components and the gradient for tuning magnet reference angle are given in Table \ref{T3} in the case of the prototype.

\begin{table}[h!]
\centering
 \begin{tabular}{||c c c c c||} 
  \hline
  & \textbf{RADIA} & \textbf{TOSCA} & \textbf{RC} & \textbf{SSW}  \\ [0.5ex] 
 \hline
 $G_0$ (T/m) & 164.5 & 164.4 &  &   \\ 
 \hline
 $\int B_2.dl$ (T.m)& 0.0658 & 0.06576 & 0.06324 & 0.0627  \\
 \hline
 $b_6$& 202 & 199 & 237 & 247  \\
 \hline 
 $b_{10}$& -158 & -152 & -133 & -138  \\[1ex] 
 \hline
 \end{tabular}
 \caption{Normalized first order multipoles and gradient results computed at 4 $mm$ radius for the prototype (tuning magnets at their reference position). RC: Rotating Coil, SSW: Single Stretched Wire.}
 \label{T3}
\end{table}

\begin{figure}[ht]
\centering
\includegraphics[scale=0.5]{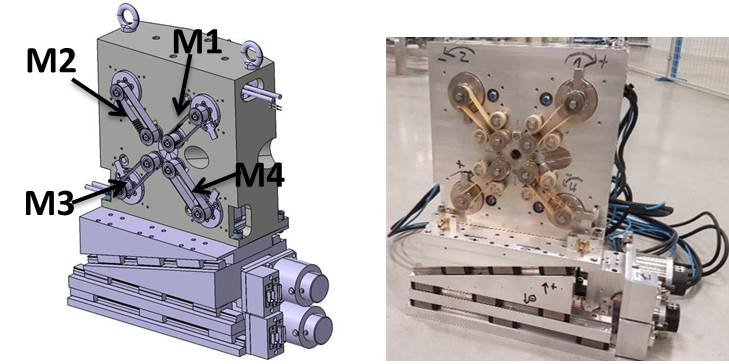}
\caption{left: Mechanical design, right: assembled QUAPEVA.}
\label{Fig4}
\end{figure}

 The chosen motors (HARMONIC DRIVE, FHA-C mini motors) have sufficient torque to counteract the magnetic forces and are very compact (48.5 x 50 x 50 mm$^3$). The magnetic system is mounted on an Aluminum frame. A non-magnetic belt transmits the rotation movement from the motor to the cylindrical magnets. The quadrupole is supported by a translation table (horizontal and vertical displacement) used to compensate any residual magnetic axis shift when varying the gradient. 
Fig. \ref{Fig4} shows the resulting mechanical design (left), and an assembled QUAPEVA on the translation table (right).

\section{\label{sec:level1} Magnetic measurements of QUAPEVA}

Three different magnetic measurements are performed to characterize the quadrupoles.
A dedicated radial rotating coil was built for the SOLEIL magnet characterization bench \cite{madur2006contribution}. A stretch-wire bench developed at ESRF \cite{lebec2012stretched}. A Pulsed-wire bench built at SOLEIL\cite{preston1992wiggler}. 

\subsection{\label{sec:level2} Gradient variability}

\begin{figure}[ht]
\centering
\includegraphics[scale=0.5]{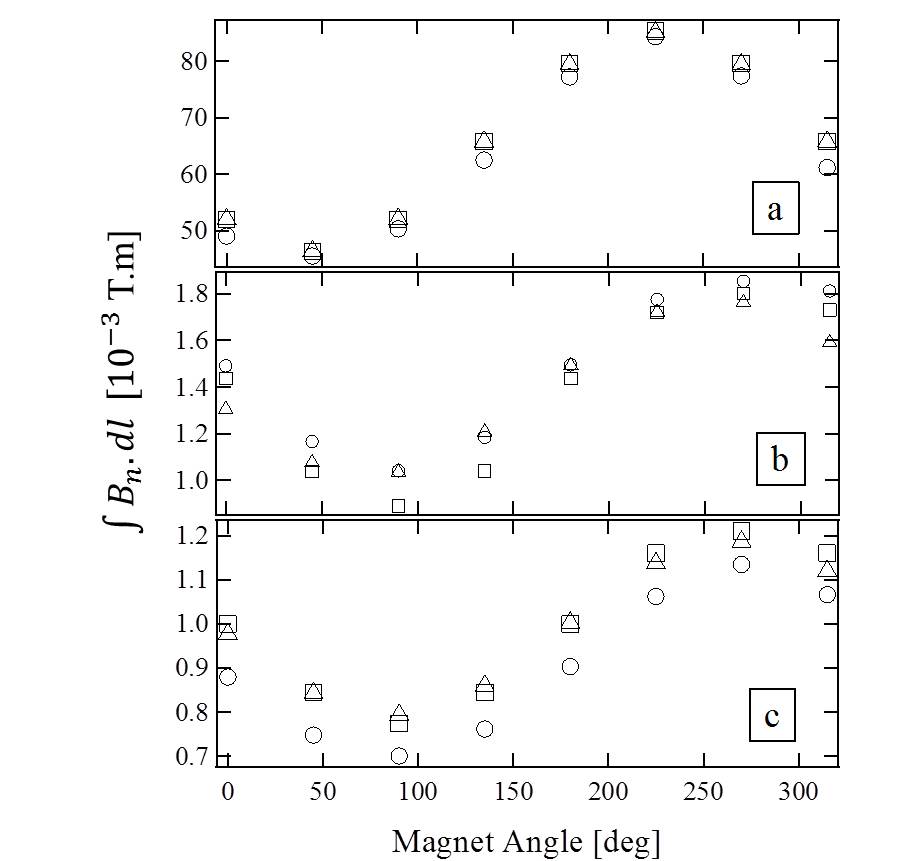}
\caption{Evolution of the main and harmonic term of the prototype QUAPEVA versus the rotation of the different cylindrical magnets. Normal multipolar terms (2,6,10) from: ($\square$) RADIA, ($\triangle$) TOSCA, ($\circ$) rotating coil measurement. a) 2$^{nd}$ (n=2), b) 6$^{th}$ (n=6), and c) 10$^{th}$ (n=10) multipolar term.}
\label{Fig5}
\end{figure}

Considering that the field in the gap is a linear combination of the field given by the main magnets and the tuning ones, a behavior model can be built. The main multipole $B_2$ becomes : 
$B_2=B_2^0+\sum_{k=1}^{k=4}B_2^k sin(\theta^k+\phi_{B_2^k})$
where $B_2^0$ is the main magnet contribution, $B_2^k$ the contribution of the $k^{th}$ tuning magnet number, $\theta^k$ its angle and $\phi_{B_2^k}$ the multipolar phase shift. Not considering the harmonic dependance with the tuning angles of the cylinders, the tuning magnet angle for a given gradient can be computed using:
$\theta_k=sin^{-1} \Big( \frac{(B_2^{req}-B_2^0)}{4 B_2^k} \Big) - \phi_{B_2^k}$
where $B_2^{req}$ is the required normal quadrupolar term.
Applying this modeling, one can then measure the gradient change of one QUAPEVA while the different cylindrical magnets are rotated simultaneously, as shown in Fig. \ref{Fig5}-a ($G=\frac{\int B_2.dl}{Rl}$), where $R$ is the radius of the measured field region. Measurement with rotating coil and stretched wire are in good agreement and corresponds to the expectations from the RADIA and TOSCA models. Fig. \ref{Fig5}-b, c shows however that the gradient variation leads to a harmonic excursion about 20$\%$ of the average value. Table \ref{T3} compares measured values to models.


%
%

\subsection{\label{sec:level2} Magnetic center evolution}

\begin{figure}[ht]
\centering
\includegraphics[scale=0.2]{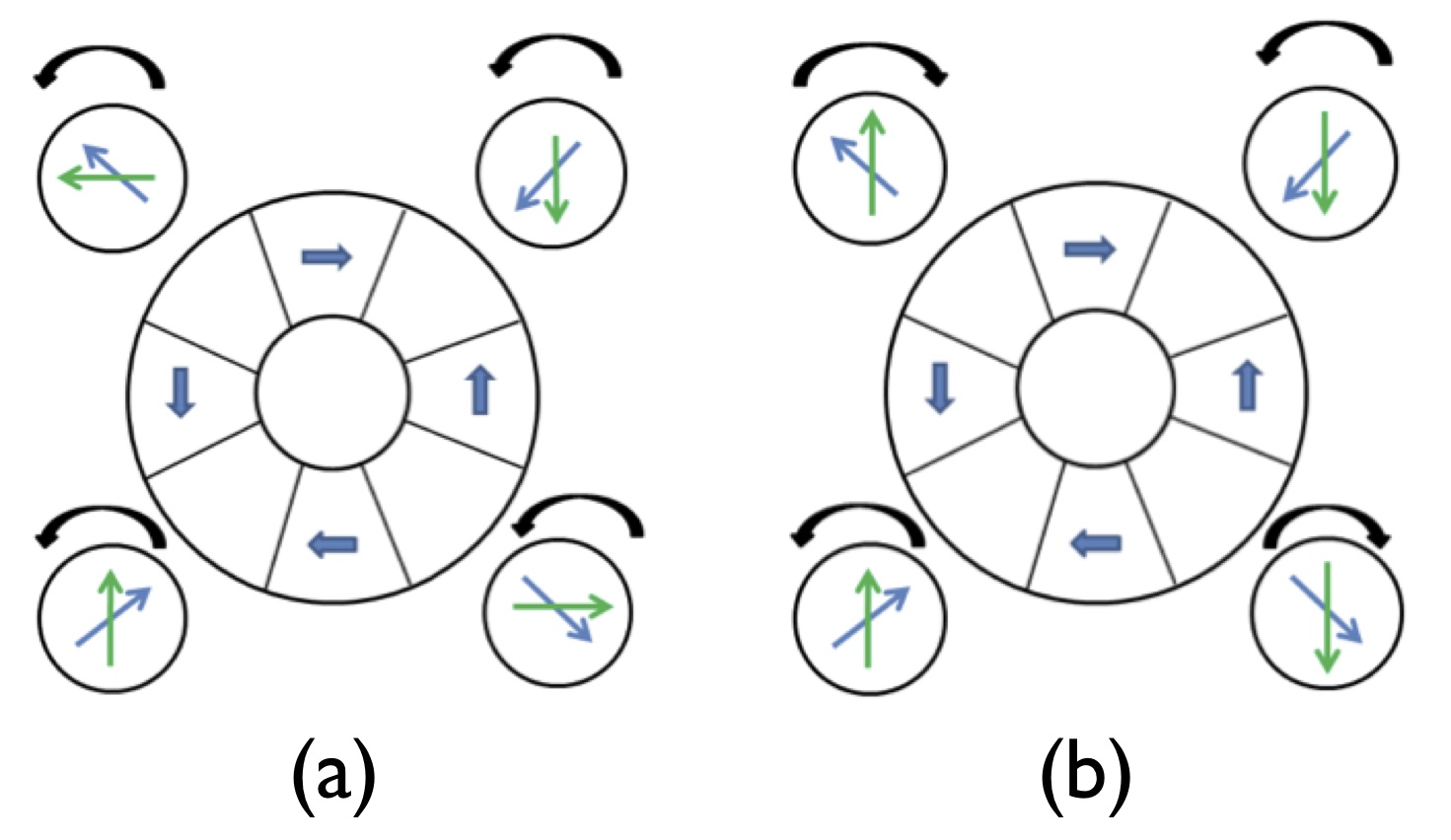}
\caption{Schemes of the starting reference position and direction of rotation for the QUAPEVA cylindrical magnets: (a) Intermediate gradient and the cylindrical magnets rotate in the same direction, (b) : Maximum gradient and the cylindrical magnets rotate in the opposite direction}
 \label{Fig6}
\end{figure}

\begin{figure}[ht]
\centering
\includegraphics[scale=0.3]{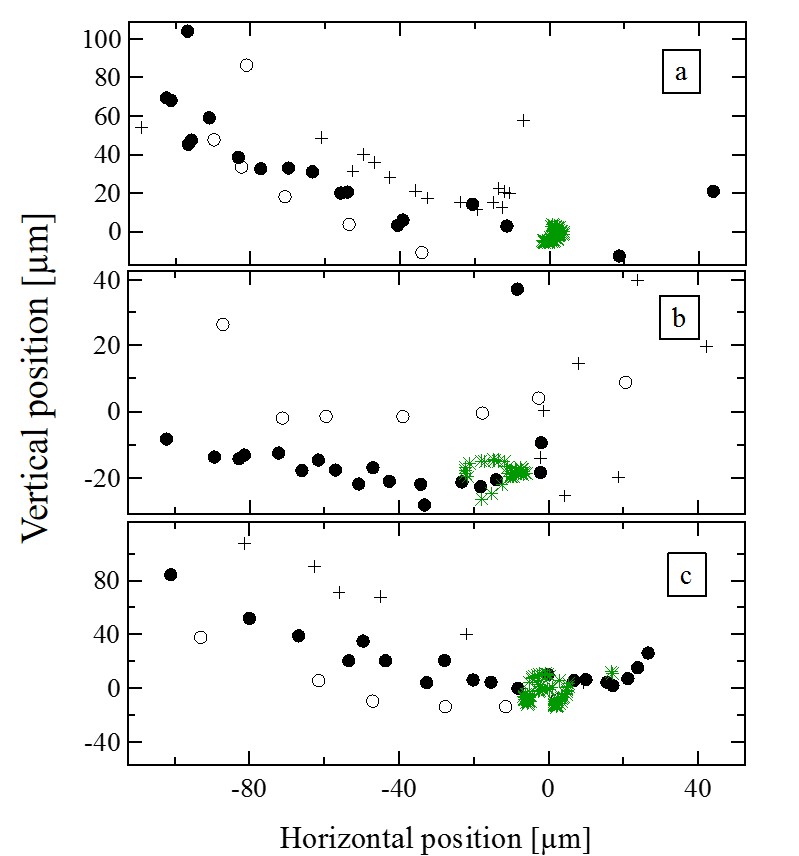}
\caption{Magnetic center evolution measurement in Fig. \ref{Fig6}-a case: ($\bullet$) pulsed wire, ($\circ$) rotating coil, (+) stretched wire measurements, and ({\color{green}$\star$}) for \ref{Fig6}-b case stretched wire measurements. QUAPEVA of a) 26 mm, (b) 40.7 mm, (c) 44.7 mm length.}
 \label{Fig7}
\end{figure}

Starting from a first reference position where the tuning magnet contribution is null and by rotating the PM cylinders in the same direction (Fig. \ref{Fig6}-a), the gradient is varied, and the horizontal and vertical excursions of the magnetic center are deduced from measurements for each different gradient (see Fig. \ref{Fig7}). The change of the magnetic axis is kept within typically $100~\mu m$. With this method the magnetic center is not very stable, since the rotation of one cylinderical magnet leads to a torque change which tends to move the other magnets away from their reference position because of the mechanical slack.

A different operating scheme (Fig. \ref{Fig6}-b) enables to start with a more stable reference position for which a change of angle induces a torque variation that brings back the magnet to its initial position. Alternating the rotation direction enables to keep symmetry. The evolution of the magnetic center versus the change of the gradient when the two different methods are applied is compared in Fig. \ref{Fig7}. The new reference position and rotation direction (in green) leads to a significant reduction of the residual excursion of the magnetic axis versus the gradient, since it remains in the $20~\mu m$ range.

\subsection{\label{sec:level2} Multipoles}

\begin{figure}[!h]
\centering
\includegraphics[scale=0.5]{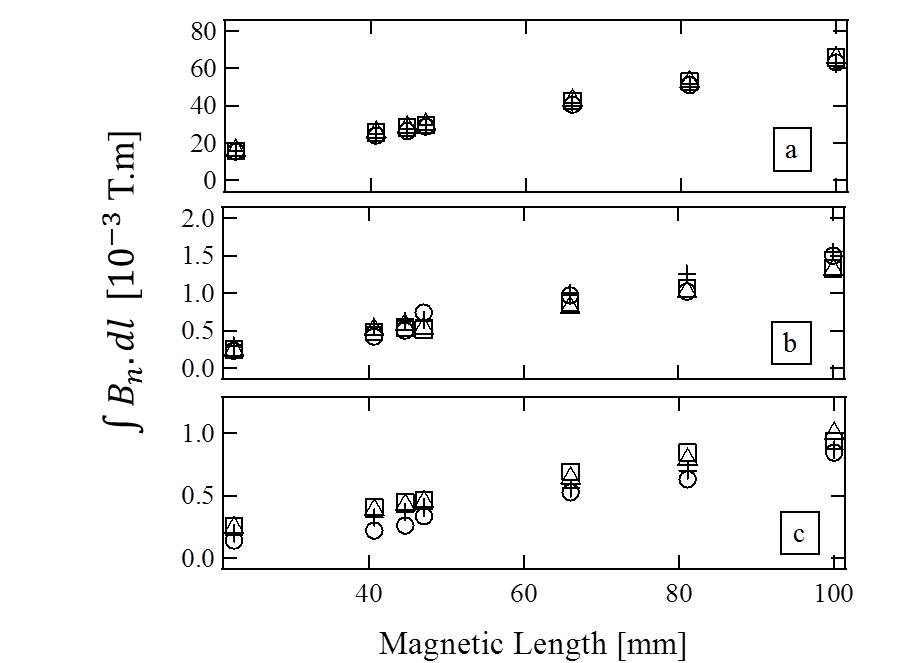}
\caption{Normal multipole terms where the cylindrical magnets are in their reference position. (a) n=2, (b) n=6, (c) n=10. ($\square$) RADIA, ($\triangle$) TOSCA, (+) stretched wire measurement, ($\circ$) rotating coil measurement.}  
\label{Fig8}
\end{figure}


Fig. \ref{Fig8} shows the systematic 2$^{nd}$, 6$^{th}$, and 10$^{th}$ multipole terms computed by the models and calculated by the measurements. Indeed they present very good agreement with a difference less than 1$\%$. The normalized skew quadrupolar term $a_2$ varies from 0.8 up to 1.8 thanks to the new position and rotation method.

\section{Application to COXINEL }

\begin{figure}[ht]
\centering
\includegraphics[scale=0.07]{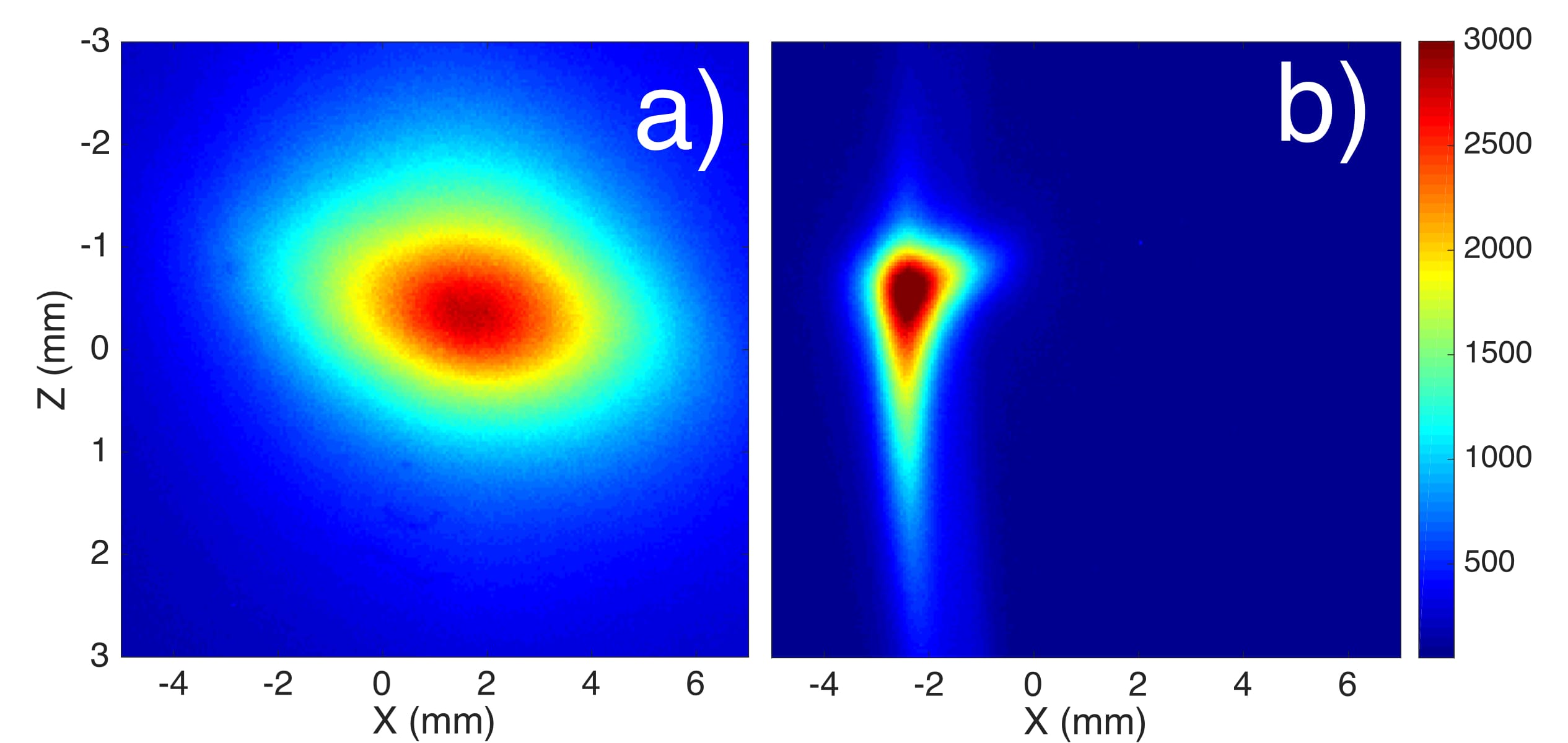}
\caption{Electron beam (Energy: 176 MeV, divergence: 5 mrad) observation on first LANEX screen at 64 cm from the source: (a) Without QUAPEVAs, (b) With QUAPEVAs}
\label{Fig9}
\end{figure}

A first triplet (26 mm, 40.7 mm, 44.7 mm mechanical length) is used for focusing the electron beam produced by laser plasma acceleration at Laboratoire d'Optique Appliqu\'ee (LOA) in view of electron qualification with a Free Electron Laser application. The results from the pulsed wire measurement have been used for QUAPEVA alignment. A beam observation on the first screen is shown in Fig. \ref{Fig9}, (a) without and (b) with the first QUAPEVA installed 5 cm away from the electron source. The large divergence of the electron beam is properly controlled and focused enabling beam 8 m long transport.

\section{Conclusion}

The design of a permanent magnet based quadrupoles of high gradient strength ($\sim$ 210 T/m) with a wide tuning range ($\sim$ 100 T/m) have been presented. These results are comforted by different magnetic measurements. The residual excursion of the magnetic center has been limited to a $\pm 10 ~\mu m$ range thanks to a particular reference position choice and rotation direction of the cylindrical magnets.
Three QUAPEVAs have been installed successively at COXINEL beam line, and are able to achieve good focusing with a highly divergent large energy spread beam.
The gradient can be enhanced $\sim 30\%$ by integrating a cooling system\cite{BenabderrahmaneNIM2012} at liquid nitrogen temperature with $Pr_2Fe_{14}B$ PMs \cite{hiroyoshi1987high}.
Besides, a design with a hyperbolic shape enables to reduce the multipole content in compromising on the gradient variability, is of great interest for low emittance storage rings.

\section{Acknowledgments}

The authors thank the European Research Council (advanced grant COXINEL - 340015), the Fondation de la Coop\'eration Scientifique / Triangle de la Physique (QUAPEVA - 2012-058T), the COXINEL team, and LOA team led by V. Malka.

\end{document}